\documentclass[12pt,a4paper]{article}
\usepackage{anziamjedraft}

\usepackage{amssymb,amsmath}
\usepackage{t1enc}
\usepackage[latin1]{inputenc}
\usepackage[german,english]{babel}
\usepackage{amsfonts}
\usepackage[all]{xy}
\usepackage{graphicx}
\usepackage{color}
\usepackage{tikz}
\usetikzlibrary{arrows,decorations.pathmorphing,backgrounds,positioning,fit,petri}

\usepackage{algorithm2e}
\restylealgo{boxruled}

\newtheorem{defi}{Definition}[section]
\newtheorem{prop}[defi]{Proposition}

\title{Non-associative key establishment protocols and their implementation}

\author{A.~G. Kalka}
\address{Department of Mathematics, Bar-Ilan University, 
Ramat Gan~52900, \textsc{Israel}.}
\mailto{arkadius.kalka@rub.de}
\http{//homepage.ruhr-uni-bochum.de/arkadius.kalka/}

\author{M. Teicher}
\address{Department of Mathematics, Bar-Ilan University, 
Ramat Gan~52900, \textsc{Israel}.}
\mailto{teicher@math.biu.ac.il}

\date{18 December 2013}

\begin{document}

\maketitle

\begin{abstract}
We provide implementation details for non-associative key establishment protocols.
In particular, we describe the implementation of non-associative key establishment protocols for all left self-distributive and all
mutually left distributive systems.
\end{abstract}

\tableofcontents

\section{Introduction} \label{sec:intro}

Currently public-key cryptography still relies mainly on a few number-theoretic problems which
remain still unbroken. Nevertheless, after the advent of quantum computers, systems like RSA, Diffie-
Hellman and ECC will be broken easily ~\cite{Sh97}.
Under the label Post Quantum Cryptography, there have been several efforts to develop new cryptographic
primitives which may also serve for the post quantum computer era. One approach became later known as
\emph{non-commutative cryptography} where the commutative groups and rings involved in number-theoretic
problems are replaced by non-commutative structures, and we consider computational problems therein ~\cite{MSU11}.
The scope of non-commutative cryptography was broadened in ~\cite{Ka07,Ka12} as we go beyond non-commutative, associative
binary oparations. 
We utilize non-associative binary operations, i.e. magmas, thus hoping to establish
\emph{non-associative public-key cryptography}. Here we focus on key establishment protocols (KEPs) as cryptographic
primitives, because they are the most important and the hardest to construct.
In particular, the seminal Anshel-Anshel-Goldfeld (AAG) KEP for monoids and groups ~\cite{AAG99} was generalized to
a general AAG-KEP for magmas in ~\cite{Ka07,Ka12} which emphasize the important and integrating role of the AAG protocol in
non-commutative and commutative cryptography.
Left self-distributive (LD) systems (and their generalizations) naturally emerge as possible non-associative
platform structures for this AAG-KEP for magmas. Non-associative key establishment protocols
for all LD-, multi-LD-, and other left distributive systems were introduced in ~\cite{KT13a,KT13b}.
Braid groups (and their finite quotients), matrix groups and Laver tables as natural platform LD-structures were discussed in 
~\cite{Ka07,Ka12, KT13a,KT13b}.
The purpose of this paper is to provide details how our non-associative KEPs can be implemented for all the systems
given in ~\cite{KT13a,KT13b}. We hope this will encourage cryptanalytic examination of these new and innovative non-associative
KEPs. \par
\emph{Outline}. In section ~\ref{sec:LD} we provide examples of LD-systems and  mutually left distributive systems.
Section ~\ref{sec:KEP} describes the most improved non-associative KEP (for all  mutually left distributive systems).
It contains all other KEPs from ~\cite{KT13a, KT13b} as special cases. Finally, section ~\ref{sec:Impl} provides implementation details
and pseudo-code.

\section{LD-systems and their generalizations}
\label{sec:LD}

\begin{defi} 
(1) An \emph{left self-distributive (LD) system} $(S,*)$ is a set $S$ equipped with a binary operation $*$ on $S$ which satisfies the 
\emph{left self-distributivity law} 
\[ x*(y*z)=(x*y)*(x*z) \quad {\rm for} \,\, {\rm all} \,\, x,y,z\in S. \] 
(2) Let $I$ be an index set. A \emph{multi-LD-system}
 $(S,(*_i)_{i\in I})$ is a set $S$ equipped with a family of binary operations $(*_i)_{i\in I}$ on $S$ such that 
\[ x*_i(y*_jz)=(x*_iy)*_j(x*_iz) \quad {\rm for} \,\, {\rm all} \,\, x,y,z\in S \]
is satisfied for every $i,j$ in $I$. Especially, it holds for $i=j$, i.e., $(S,*_i)$ is an LD-system. If $|I|=2$ then we call $S$ a \emph{bi-LD-system}.
\par (3) A \emph{mutually left distributive system} $(S, *_a, *_b)$ is a set $S$ equipped with two binary operations $*_a, *_b$ on $S$ such that 
\[ x*_a(y*_bz)=(x*_ay)*_b(x*_az) \quad  x*_b(y*_az)=(x*_by)*_a(x*_bz)  \quad {\rm for} \,\, {\rm all} \,\, x,y,z\in S. \] 
\end{defi}

More vaguely, we will also use the terms \emph{partial multi-LD-system} and simply \emph{left distributive system} if the laws of a multi-LD-system
are only fulfilled for special subsets of $S$ or if only some of these (left) distributive laws are satisfied. 
A mutually left distributive system $(L, *_a, *_b)$ is only a partial bi-LD-system. 
The left selfdistributivity laws need not hold, i.e., $(L,*_a)$ and $(L, *_b)$ are in general not LD-systems. 
We list examples of LD-systems, multi-LD-systems and mutually left distributive systems. 
More details can be found in ~\cite{De00, De06, Ka12, KT13a, KT13b}.

\emph{Conjugacy.} A classical example of an LD-system is $(G,*)$ where $G$ is a group equipped with the conjugacy operation $x*y=x^{-1}yx$ (or $x*^{\rm rev}y=xyx^{-1}$).

\emph{Laver tables.} Finite groups equipped with the conjugacy operation are not the only finite LD-systems. Indeed, the socalled \emph{Laver tables} provide the classical example for finite LD-systems. 
There exists for each $n\in \mathbb{N}$ an unique LD-system $L_n=(\{ 1, 2, \ldots , 2^n \},*)$ with $k*1=k+1$.
The values for $k*l$ with $l\ne 1$ can be computed by induction using the left self-distributive law.
Laver tables are also described in ~\cite{De00}. 

\emph{LD-conjugacy.}
Let $G$ be a group, and $f\in End(G)$. Set $x*_fy=f(x^{-1}y)x$, then $(G,*_f)$ is an LD-system. 

\emph{Shifted conjugacy.}
Consider the braid group on infinitely many strands
\[ B_{\infty }= \langle \{\sigma _i \}_{i\ge 1} \mid \sigma _i\sigma _j=\sigma _j\sigma _i \,\,{\rm for} \,\, |i-j|\ge 2, \,\, 
\sigma _i\sigma _j\sigma _i=\sigma _j\sigma _i\sigma _j \,\, {\rm for} \,\, |i-j|=1\rangle  \] 
where inside $\sigma _i$ the $(i+1)$-th strand crosses over the $i$-th strand. 
The {\it shift map} $\partial : B_{\infty } \longrightarrow B_{\infty }$ defined by $\sigma _i \mapsto \sigma _{i+1}$ for all $i\ge 1$
is an injective endomorphism. Then
$B_{\infty }$ equipped with the {\rm shifted conjugacy} operations $*$, $\bar{*}$ defined by
$x*y=\partial x^{-1}\cdot \sigma _1 \cdot \partial y \cdot x$ and $x\, \bar{*}\, y=\partial x^{-1}\cdot \sigma _1^{-1} \cdot \partial y  \cdot x$
is a bi-LD-system. In particular, $(B_{\infty },*)$ is an LD-system. 
 
\emph{Generalized shifted conjugacy in braid groups.}
Let, for $n\ge 2$,  $\delta _n=\sigma _{n-1}\cdots \sigma _2\sigma _1$. For $p, q \ge 1$, we set
$\tau _{p,q}=\delta _{p+1}\partial (\delta _{p+1})\cdots \partial ^{q-1}(\delta _{p+1})$. 

\begin{prop} \label{abc}
$(B_{\infty }, *_1, *_2)$ with binary operations $x*_iy=\partial ^p(x^{-1})a_i\partial ^p(y)x$ ($i=1,2$) with $a_1=a_1'\tau _{p,p}^{\pm 1}a_1''$, 
$a_2=a_2'\tau _{p,p}^{\pm 1}a_2''$ for some $a_1',a_1'',a_2',a_2''\in B_p$ is a mutually left distributive system 
if and only if $[a_1',a_2'']=[a_2',a_1'']=[a_1',a_2']=1$.
(Note that $[a_1',a_1'']$, $[a_2',a_2'']$ and $[a_1'',a_2'']$ may be nontrivial. If, in addition $[a_1',a_1'']=[a_2',a_2'']=1$ holds, then $(B_{\infty }, *_1, *_2)$ is a bi-LD-system.) 
\end{prop}

\emph{Symmetric conjugacy.} 
For a group $G$, there exists yet another LD-operation. $(G,\circ )$ is an LD-system with $x\circ y=xy^{-1}x$. 

\emph{$f$-symmetric conjugacy.}
Let $G$ be a group, and $f\in End(G)$ an endomorphism that is also a projector ($f^2=f$).
Then $(G, \circ _f)$, defined by  $x\circ _f y=f(xy^{-1})x$ is an LD-system.

\section{Non-associative KEPs for mutually left distributive systems}
\label{sec:KEP}

Here we describe a KEP that works for all mutually left distributive systems, in particular all bi-LD-systems (and all LD-systems).
Consider a set $L$ equipped with a pool of binary operations $O_A \cup O_B$ ($O_A$ and $O_B$ non-empty) s.t.
the operations in $O_A$ are distributive over those in $O_B$ and vice versa, i.e. the following holds
for all $x,y,z\in L$, $*_{\alpha } \in O_A$ and $*_{\beta }\in O_B$.
\begin{eqnarray}
  x*_{\alpha }(y*_{\beta }z)&=&(x*_{\alpha }y)*_{\beta }(x*_{\alpha }z), \,\, {\rm and}    \label{abLD} \\
  x*_{\beta }(y*_{\alpha }z)&=&(x*_{\beta }y)*_{\alpha }(x*_{\beta }z).                    \label{baLD}
\end{eqnarray}
Then $(L, *_{\alpha }, *_{\beta })$ is a mutually left distributive system for all $(*_{\alpha },*_{\beta })\in O_A \times O_B$.
Note that, if $O_A \cap O_B \ne \emptyset$, then $(L, O_A \cap O_B)$ is a multi-LD-system. 
\par
\begin{figure} \label{LDTree}
  \caption{$(s_3 *_{\alpha _1} (( s_3 *_{\alpha _4} (s_1 *_{\alpha _1} s_2)) *_{\alpha _2} s_1 )) *_{\alpha _1} (( s_2 *_{\alpha _2} s_3) *_{\alpha _3} s_2) \in S_A$}
\begin{center}
\begin{tikzpicture} [scale = 0.7]
   \node (r1) at (0,0) [circle,inner sep=1pt,draw=black!50,fill=black!20]{}; \node[below] at (r1.south) {$s_3$};  
   \node (r2) at (2,0) [circle,inner sep=1pt,draw=black!50,fill=black!20]{}; \node[below] at (r2.south) {$s_3$};  
   \node (r3) at (4,0) [circle,inner sep=1pt,draw=black!50,fill=black!20]{}; \node[below] at (r3.south) {$s_1$};  
   \node (r4) at (6,0) [circle,inner sep=1pt,draw=black!50,fill=black!20]{}; \node[below] at (r4.south) {$s_2$};  
   \node (r5) at (8,0) [circle,inner sep=1pt,draw=black!50,fill=black!20]{}; \node[below] at (r5.south) {$s_1$}; 
   \node (r6) at (10,0) [circle,inner sep=1pt,draw=black!50,fill=black!20]{}; \node[below] at (r6.south) {$s_2$};  
   \node (r7) at (12,0) [circle,inner sep=1pt,draw=black!50,fill=black!20]{}; \node[below] at (r7.south) {$s_3$};  
   \node (r8) at (14,0) [circle,inner sep=1pt,draw=black!50,fill=black!20]{}; \node[below] at (r8.south) {$s_2$};  

   \node (i34) at (5,1) [circle,inner sep=1pt,draw=black!50,fill=black!20]{}
          edge[thick] (r3)   edge[thick] (r4);     \node[below] at (i34.south) {$*_{\alpha _1}$};  
   \node (i24) at (4,2) [circle,inner sep=1pt,draw=black!50,fill=black!20]{}
          edge[thick] (r2)   edge[thick] (i34);     \node[below] at (i24.south) {$*_{\alpha _4}$};  
   \node (i25) at (5,3) [circle,inner sep=1pt,draw=black!50,fill=black!20]{}
          edge[thick] (i24)   edge[thick] (r5);     \node[below] at (i25.south) {$*_{\alpha _2}$};  
   \node (i15) at (4,4) [circle,inner sep=1pt,draw=black!50,fill=black!20]{}
          edge[thick] (r1)   edge[thick] (i25);   \node[below] at (i15.south) {$*_{\alpha _1}$}; 
   \node (i67) at (11,1) [circle,inner sep=1pt,draw=black!50,fill=black!20]{}
          edge[thick] (r6)   edge[thick] (r7);     \node[below] at (i67.south) {$*_{\alpha _2}$};  
   \node (i68) at (12,2) [circle,inner sep=1pt,draw=black!50,fill=black!20]{}
          edge[thick] (i67)   edge[thick] (r8);     \node[below] at (i68.south) {$*_{\alpha _3}$};  
   \node (i18) at (7,7) [circle,inner sep=1pt,draw=black!50,fill=black!20]{}
          edge[thick] (i15)   edge[thick] (i68);   \node[below] at (i18.south) {$*_{\alpha _1}$}; 
\end{tikzpicture}
\end{center}
\end{figure}
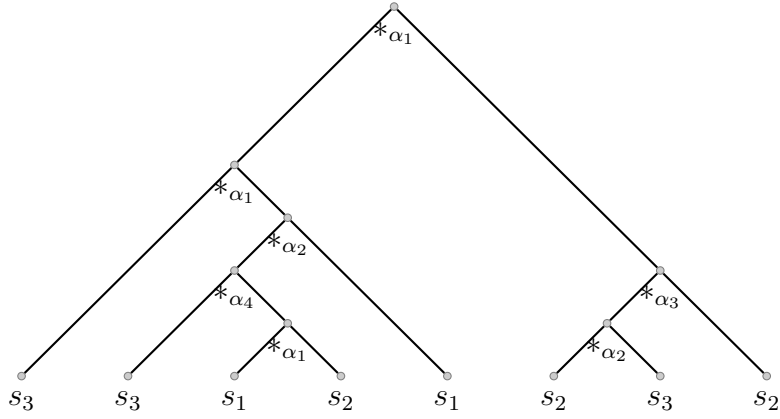

 Let $s_1, \ldots , s_m, t_1, \ldots , t_n\in L$ be some public elements. We denote
$S_A=\langle s_1, \cdots , s_m \rangle _{O_A}$ and $S_B=\langle t_1, \cdots , t_n \rangle _{O_B}$, two submagmas of $(L,O_A\cup O_B)$.
For example, an element $y$ of $S_A$ can be described by a planar rooted binary
tree $T$ whose $k$ leaves are labelled by these other elements $r_1,\ldots ,r_k$ with $r_i \in \{s_i\}_{i\le m}$.
Here the tree contains further information, namely to each internal vertex we assign a binary operation $*_i \in O_A$.
We use the notation $y=T_{O_A}(r_1,\ldots ,r_k)$. 
The subscript $O_A $ tells us that the grafting of subtrees of $T$ corresponds to the operation $*_i\in O_A$.
Consider, for example, the element $y=(s_3 *_{\alpha _1} (( s_3 *_{\alpha _4} (s_1 *_{\alpha _1} s_2)) *_{\alpha _2} s_1 )) *_{\alpha _1} (( s_2 *_{\alpha _2} s_3) *_{\alpha _3} s_2)$. The corresponding
labelled planar rooted binary tree $T$ is displayed in the Figure ~\ref{LDTree}.

Let $*_{\alpha }\in O_A$ and $*_{\beta }\in O_B$. By induction over the tree depth, it is easy to show that, for all elements $e, e_1, \ldots , e_l \in (L, O_A \cup O_B)$ and all planar rooted binary trees $T$ with $l$ leaves, the following equations hold.
\begin{eqnarray}
e*_{\alpha }T_{O_B}(e_1, \ldots , e_l)&=&T_{O_B}(e*_{\alpha }e_1, \ldots , e*_{\alpha }e_l),   \\
e*_{\beta }T_{O_A}(e_1, \ldots , e_l)&=&T_{O_A}(e*_{\beta }e_1, \ldots , e*_{\beta }e_l).
\end{eqnarray}

\begin{prop} \label{itLDendoMutual} (See Proposition 4.1 in ~\cite{KT13b}.)
Consider $(L,O_A \cup O_B)$ such that $(L, *_A, *_B)$ is a mutually left distributive system for all $(*_A, *_B)\in O_A \times O_B$, 
and let $k \in \mathbb{N}$.
Then, for all $x=(x_1, \ldots , x_k) \in L^k$, $o_A=(*_{A_1}, \ldots ,*_{A_k}) \in O_A^k$, and $o_B=(*_{B_1}, \ldots ,*_{B_k}) \in O_B^k$, 
the iterated left multiplication maps
\begin{eqnarray*} 
\phi _{(x,o_A)}: && y \mapsto x_k*_{A_k}(x_{k-1}*_{A_{k-1}} \cdots *_{A_3}(x_2 *_{A_2}(x_1*_{A_1}y)) \cdots ) \,\, {\rm and}  \\
\phi _{(x,o_B)}: && y \mapsto x_k*_{B_k}(x_{k-1}*_{B_{k-1}} \cdots *_{B_3}(x_2 *_{B_2}(x_1*_{B_1}y)) \cdots )
\end{eqnarray*}
define a magma endomorphisms of $(L, O_B)$ and $(L, O_A)$, respectively.
\end{prop}

In particular, the following equations hold for all $k,l \in \mathbb{N}$, $a, b \in L^k$, $o_A\in O_A^k$, $o_B\in O_B^k$,
$e, e_1, \ldots , e_l \in L$ and all planar rooted binary trees $T$ with $l$ leaves.
\begin{eqnarray}
\phi _{(a,o_A)} (T_{O_B}(e_1, \ldots , e_l))&=&T_{O_B}(\phi _{(a,o_A)}(e_1), \ldots , \phi _{(a,o_A)}(e_l)),   \\
\phi _{(b,o_B)} (T_{O_A}(e_1, \ldots , e_l))&=&T_{O_A}(\phi _{(b,o_B)}(e_1), \ldots , \phi _{(b,o_B)}(e_l))
\end{eqnarray}

Now, we are going to describe a KEP that applies to any system $(L,O_A\cup O_B)$ as described above.
We have two subsets of public elements $\{ s_1, \cdots , s_m \}$ and $\{t_1, \cdots , t_n \}$ of $L$.
Also, recall that $S_A=\langle s_1, \cdots , s_m \rangle _{O_A}$ and $S_B=\langle t_1, \cdots , t_n \rangle _{O_B}$.
Alice and Bob perform the following protocol steps.

\begin{description}
\item[{\bf Protocol}] {\sc Key establishment for the partial multi-LD-system} \par
$(L,O_A\cup O_B)$.
\item[{\rm 1}] Alice generates her secret key $(a_0,a, o_A) \in S_A \times L^{k_A} \times O_A^{k_A}$, and Bob chooses his secret key 
   $(b, o_B)\in S_B^{k_B} \times O_B^{k_B}$. Denote $o_A=(*_{A_1}, \ldots , *_{A_{k_A}})$ and $o_B=(*_{B_1}, \ldots , *_{B_{k_B}})$, 
 then Alice's and Bob's secret magma morphisms $\alpha $ and $\beta $ are given by
\begin{eqnarray*} 
\alpha (y)&=&a_{k_A}*_{A_{k_A}}(a_{k_A-1}*_{A_{k_A-1}} \cdots *_{A_3}(a_2 *_{A_2}(a_1*_{A_1}y)) \cdots ) \quad {\rm and} \\ 
\beta (y)&=&b_{k_B}*_{B_{k_B}}(b_{k_B-1}*_{B_{k_B-1}} \cdots *_{B_3}(b_2 *_{B_2}(b_1*_{B_1}y)) \cdots ),
\end{eqnarray*}
respectively.
\item[{\rm 2}] $(\alpha (t_i))_{1\le i \le n} \in L^n, p_0= \alpha (a_0) \in L$, and sends them to Bob. 
Bob computes the vector $(\beta (s_j))_{1\le j \le m} \in L^m$, and sends it to Alice. 
\item[{\rm 3}] Alice, knowing $a_0=T_{O_A}(r_1, \ldots , r_l)$ with $r_i\in \{s_1,\ldots ,s_m\}$, computes from the received message
\[ T_{O_A}( \beta (r_1), \ldots , \beta (r_l))=\beta (T_{O_A}(r_1, \ldots , r_l))=\beta (a_0). \]
And Bob, knowing for all $1\le j \le k_B$, $b_j=T^{(j)}_{O_B}(u_{j,1}, \ldots , u_{j,l_j})$ with $u_{j,i}\in \{t_1,\ldots ,t_n\} \forall i\le l_j$ for some
$l_j \in \mathbb{N}$, computes from his received message for all $1\le j \le k_B$
\[ T^{(j)}_{O_B}(\alpha (u_{j,1}), \ldots , \alpha (u_{j,l_j}))=\alpha (T^{(j)}_{O_B}(u_{j,1}, \ldots , u_{j,l_j})=\alpha (b_j). \]
\item[{\rm 4}] Alice computes $K_A=\alpha (\beta (a_0))$.
Bob gets the shared key by 
\[ K_B:=\alpha (b_{k_B})*( \alpha (b_{k_B-1})*( \cdots (\alpha (b_2)*(\alpha (b_1)*p_0)) \cdots ))\stackrel{\alpha \, {\rm homo}}{=}K_A. \]
\end{description}

\begin{figure}
  \caption{{\sc KEP for the partial multi-LD-system} $(L,O_A\cup O_B)$.}
\begin{center} 
 \begin{tikzpicture}
 \node[red]  (Alice) at ( 0,0) [circle,draw=black!50,fill=black!20]{Alice};
 \node[blue]  (Bob) at ( 8.5,0) [circle,draw=black!50,fill=black!20]{Bob}
   edge [<-, bend right=10] node[auto,swap] (pA) {$\{ \phi _{{\color{red}(a,o_A)}}({\color{green}t_i}) \}_{1\le i \le n}, \,\,\, 
   \phi _{{\color{red}(a,o_A)}}({\color{red}a_0})$} (Alice)   
   edge [->, bend left=10] node[auto] (pB) { $\{ \phi _{{\color{red}(b,o_B)}}({\color{green}s_j}) \}_{1\le j \le m} $} (Alice) ;
 \node (skA) [below] at (Alice.south) [rectangle,draw=red!50,fill=red!20]{${\color{red}a_0}\in S_A, {\color{red}a} \in L^{{\color{red}k_A}}, {\color{red}o_A} \in O_A^{{\color{red}k_A}}$};
 \node (skB) [below] at (Bob.south) [rectangle,draw=red!50,fill=red!20]{${\color{red}b}\in S_B^{{\color{red}k_B}}, {\color{red}o_B} \in O_B^{{\color{red}k_B}}$};
 \begin{pgfonlayer}{background}
  \node [fill=green!20, rounded corners, fit=(Alice) (Bob) (skA) (skB) (pA) (pB)] {};
 \end{pgfonlayer} 
 \end{tikzpicture}
\end{center}
\end{figure}
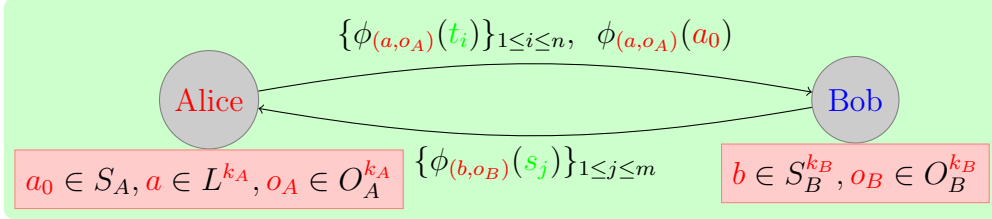

Here the operation vectors $o_A \in O_A^{k_A}$ and $o_B \in O_B^{k_B}$ are part of Alice's and Bob's private keys.
Also explicit expressions of $a_0\in S_A$ and all $b_i \in S_B$ as treewords $T, T^{(i)}$ (for all $1\le i \le k_B$) 
are also parts of the private keys - though we did not mention it explicitly in step 1 of the protocols. But here $T_{O_A}$ and $T'_{O_B}$ also contain all the information about the grafting operations (in $O_A$ or $O_B$, respectively) at the internal vertices of $T$, $T^{(1)}, \ldots , T^{(k_B)}$.

\section{Implementation}
\label{sec:Impl}

\emph{Planar rooted binary trees} We need some efficient way to encode the planar rooted binary tree which determines the bracket structure of an element given as product of other elements. Let $PBT_n$ denote the set of planar rooted binary trees (also known as \emph{full binary trees}) with
$n$ internal nodes (and $n+1$ leaves), then  $|PBT_n|=Cat(n)$ where $Cat(n)=\frac{1}{n+1}{2n \choose n}$ denotes the $n$-th Catalan number.
There exists a rich variety of other Catalan sets with well understood bijections between them, e.g., diagonal avoiding paths (aka mountain ranges), 
polygon triangulations, Dyck words, planar rooted trees (not only binary) and non-crossing partitions. 
We use the following succinct representation for Catalan sets taken from ~\cite{CM09}. 
Denote $[n]=\{1, \ldots , n\}$. To each $T\in PBT_n$ we associate a vector (array) $T \in [n]^n$ such that
$T[i] \le T[j]$ for $i<j$ and $T[i]\le i$ for all $i \in [n]$. By abuse of notation we call the set of such vectors in $[n]^n$ also $PBT_n$.

\begin{algorithm}{{\bf function} {\sf EvaluateTree}}\;
\KwIn{$(T, o, (e_1,\ldots , e_{n+1})) \in PBT_n \times O^n \times L^{n+1}$.}
\KwOut{$e=T_o(e_1, \ldots , e_{n+1})$}
 \For{$j:=n$ {\bf to} 1 {\bf by} -1}{
   $pos:=T[j]$\;
   $Seq[pos]:=Seq[pos] *_{o[pos]} Seq[pos +1]$\;
   ${\sf Remove}(\,\tilde{}\,Seq, pos +1)$;  \, \, ${\sf Remove}(\,\tilde{}\,T, pos)$;   \, \, ${\sf Remove}(\,\tilde{}\,o, pos)$\;    
 }
 {\bf return} $Seq[1]$\;
\end{algorithm}

Let $L$ be a magma and $O$ be a set of binary operations on $L$. Given a vector of operations $o=(*_{o[1]}, \ldots , *_{o[n]})\in O^n$ and 
a sequence of leave elements 
$(e_1, \ldots , e_{n+1}) \in L^{n+1}$, then the function {\sf EvaluateTree} evaluates the product of $e_1, \ldots , e_{n+1}$ where the bracket structure is given by the
tree $T$ and the operations on the internal vertices of $T$ are given by $o$.
For example, the tree in Figure 1 is given by $T=[1,1,2,2,3,6,6]$ and $o=(*_{\alpha _2}, *_{\alpha _3}, *_{\alpha _1}, *_{\alpha _4}, *_{\alpha _2}, *_{\alpha _1}, *_{\alpha _1})$. 
\par
\emph{Protocol implementation.} Now, let $(L,O_A,O_B)$ be as described in the KEP.  
We fix some distributions on $L$, $O_A$ and $O_B$, so that we may generate random elements from these sets (according to these distributions).
Given $m_a, m_B \in \mathbb{N}$, Alice and Bob first choose 
random vectors $\mathcal{G}_A=(s_1, \ldots , s_{m_A}) \in L^{m_A}$ and $\mathcal{G}_B =(t_1, \ldots , t_{m_B})\in L^{m_B}$ which determine
the public submagmas $S_A=\langle \mathcal{G}_A \rangle _{O_A}$ and $S_B=\langle \mathcal{G}_B \rangle _{O_B}$, respectively.
Then Alice and Bob generate their secret, public and shared keys as described in the following functions.
\par The KEPs were implemented using MAGMA \cite{MAG} which also contains an implementation of braid groups following \cite{CK+01}. 

\begin{algorithm}{{\bf function} {\sf GeneratePrivateKeyAlice}}\;
\KwIn{$\mathcal{G}_A \in L^{m_A}$.}
\KwOut{$(Ia_0, Ta_0 , oa_0, a_0, a, oA) \in [m_A]^{n_{a_0}+1} \times PBT_{n_{a_0}} \times O_A^{n_{a_0}} \times L \times L^{k_A} \times O_A^{k_A}$}
  $Ia_0 \leftarrow {\sf Random}([m_A]^{n_{a_0}})$\;
  \lFor{$i:=1$ {\bf to} $n_{a_0}+1$}{  $Seqa_0[i]:=\mathcal{G}_A[Ia_0[i]]$\; }
  $Ta_0 \leftarrow {\sf Random}(PBT_{n_{a_0}})$; \,\,   $oa_0 \leftarrow {\sf Random}(O_A^{n_{a_0}})$\;
  $a_0:={\sf EvaluateTree}(Ta_0, oa_0, Seqa_0)$\;
  $a \leftarrow {\sf Random}(L^{k_A})$; \,\,   $oA \leftarrow {\sf Random}(O_A^{k_A})$\;
 {\bf return} $(Ia_0, Ta_0 , oa_0, a_0, a, oA)$\;

\vspace{0.4cm}

{\bf function} {\sf GeneratePrivateKeyBob}\;
\KwIn{$\mathcal{G}_B \in L^{m_B}$.}
\KwOut{$(Ib, Tb , ob, b, oB) \in ([m_B]^{n_b+1})^{k_B} \times (PBT_{n_b})^{k_B} \times (O_A^{n_b})^{k_B} \times L^{k_B} \times O_B^{k_B}$}
 \For{$k:=1$ {\bf to} $k_B$}{
    $Ib[k] \leftarrow {\sf Random}([m_B]^{n_b})$\;
    \lFor{$i:=1$ {\bf to} $n_b+1$}{  $Seqb[k][i]:=\mathcal{G}_A[Ia_0[i]]$\; }
    $Tb[k] \leftarrow {\sf Random}(PBT_{n_b})$; \,\,   $ob \leftarrow {\sf Random}(O_B^{n_b})$\;
    $b[k]:={\sf EvaluateTree}(Tb[k], ob[k], Seqb[k])$\;
 }    
 $oB \leftarrow {\sf Random}(O_B^{k_B})$\;
 {\bf return} $(Ib, Tb , ob, b, oB)$\;

\vspace{0.4cm}

{\bf function} {\sf GeneratePublicKeyAlice}\;
\KwIn{$(a, oA, a_0, \mathcal{G}_B) \in L^{k_A} \times O_A^{k_A} \times L \times L^{m_B}$.}
\KwOut{$(p_A, p_0) \in L^{m_A} \times L$}
 \For{$k:=1$ {\bf to} $m_A$}{
    $p_A[k]:=\mathcal{G}_B[k]$\;
    \lFor{$i:=1$ {\bf to} $k_A$}{  $p_A[k]:=a[i] *_{oA[i]} p_A[k]$\; }
 }    
 $p_0:=a_0$\;
 \lFor{$i:=1$ {\bf to} $k_A$}{  $p_0[k]:=a[i] *_{oA[i]} p0[k]$\; } 
 {\bf return} $(p_A, p_0)$\;

\vspace{0.4cm}

{\bf function} {\sf GeneratePublicKeyBob}\;
\KwIn{$(b, oB, \mathcal{G}_A) \in L^{k_B} \times O_B^{k_B} \times L^{m_A}$.}
\KwOut{$p_B \in L^{m_B}$}
 \For{$k:=1$ {\bf to} $m_B$}{
    $p_B[k]:=\mathcal{G}_A[k]$\;
    \lFor{$i:=1$ {\bf to} $k_B$}{  $p_B[k]:=b[i] *_{oB[i]} p_B[k]$\; }
 }     
 {\bf return} $p_B$\;
\end{algorithm}

\begin{algorithm}{{\bf function} {\sf GenerateSharedKeyAlice}}\;
\KwIn{$(Ia_0, Ta_0 , oa_0, a, oA, p_B) \in [m_A]^{n_{a_0}+1} \times PBT_{n_{a_0}} \times O_A^{n_{a_0}}  \times L^{k_A} \times O_A^{k_A} \times L^{m_B}$.}
\KwOut{$K_A \in L$}
 $K_A:={\sf EvaluateTree}(Ta_0, oa_0, (p_B[Ia_0[i]])_{i\le n_{a_0}+1})$\;
 \lFor{$k:=1$ {\bf to} $k_A$}{  $K_A:=a[k] *_{oA[k]} K_A$\; }     
 {\bf return} $K_A$\;

\vspace{0.4cm}

{\bf function} {\sf GenerateSharedKeyBob}\;
\KwIn{$(Ib, Tb , ob, b, oB, p_A, p_0) \in ([m_B]^{n_b+1})^{k_B} \times (PBT_{n_b})^{k_B} \times (O_A^{n_b})^{k_B} \times L^{k_B} \times O_B^{k_B} \times L^{m_A} \times L$.}
\KwOut{$K_A \in L$}
 Initialize $lfactors:=[]$; $K_B:=p_0$\;
 \For{$k:=1$ {\bf to} $k_B$}{ $lfactors[k]:={\sf EvaluateTree}(Tb[k], ob[k], (p_A[Ib[k][i]])_{i\le n_b+1})$\; 
 $K_B:=lfactors[k] *_{oB[k]} K_B$\; 
 }  
 {\bf return} $K_A$\;
\label{GenKeys}
\end{algorithm}


\bibliographystyle{plain}
\bibliography{refs-emacsubmit.bib}

\end{document}